%% file: reengineering.tex
\documentclass[copyright]{eptcs}
\providecommand{\event}{TTC 2011} 

\usepackage{tikz}
\usetikzlibrary{block}

\usepackage{listings}
\usepackage{graphicx}

\newcommand{\origttfamily}{}
\let\origttfamily=\ttfamily
\renewcommand{\ttfamily}{\origttfamily \hyphenchar\font=`\-}

\newcommand{\name}[1]{{\sc #1}}

\newcommand{\grgennet}{\name{GrGen.NET}}
\newcommand{\grshell}{\name{GrShell}}

\lstdefinelanguage{grgen}
{morekeywords={actions,using,rule,test,pattern,replace,if,for,eval,negative,independent,node,edge,graph,modify,delete,class,model,connect,enum,abstract,const,exec,emit,alternative,multiple,iterated,optional,yield,copy,sequence,def,evalhere,emithere}, 
sensitive=true,
morecomment=[l]{//},
morecomment=[s]{/*}{*/},
morestring=[b]",
basicstyle=\ttfamily\footnotesize,
keywordstyle=\itshape
}

\lstdefinelanguage{grshell}
{morekeywords={xgrs,debug,import,dump,add,set,node,edge,group,by,hidden,outgoing,labels,off,shortinfotag,exclude,shape,color,rhomb,white},
sensitive=true,
morecomment=[l]{\#},
morecomment=[s]{/*}{*/},
morestring=[b]",
basicstyle=\ttfamily\footnotesize
}

\title{Solving the TTC 2011 Reengineering Case with GrGen.NET}
\author{Edgar Jakumeit \quad \quad Sebastian Buchwald
\institute{Karlsruhe Institute of Technology (KIT)}
\email{\phantom{~~~~~~~~~~~~~~~~~~} \quad \quad buchwald@kit.edu}
}
\def\titlerunning{Solving the TTC 2011 Reengineering Case with GrGen.NET}
\def\authorrunning{Edgar Jakumeit and Sebastian Buchwald}
\begin{document}
\maketitle

\begin{abstract}
The challenge of the Reengineering Case~\cite{programunderstandingcase} is to extract a state machine model out of the abstract syntax graph of a Java program.
The extracted state machine offers a reduced view on the full program graph and thus helps to understand the program regarding the question of interest.
We tackle this task employing the general purpose graph rewrite system GrGen.NET (\url{www.grgen.net}).
\end{abstract}

\section{What is GrGen.NET?}

\grgennet\ is an application domain neutral graph rewrite system~\cite{GrGenUserManual}, the feature highlights regarding practical relevance are:
\begin{description}\itemsep -2pt
\item[Fully Featured Meta Model:] \grgennet\ uses attributed and typed multigraphs with multiple inheritance on node/edge types. Attributes may be typed with one of several basic types, user defined enums, or generic set, map, and array types.
\item[Expressive Rules, Fast Execution:] The expressive and easy to learn rule specification language allows straightforward formulation of even complex problems, with an optimized implementation yielding high execution speed at modest memory consumption.
\item[Programmed Rule Application:] \grgennet\ supports a high-level rule application control language, Graph Rewrite Sequences (GRS), offering sequential, logical, iterative and recursive control plus variables and storages for the communication of processing locations between rules.
\item[Graphical Debugging:] \grshell, \grgennet's command line shell, offers interactive execution of rules, visualizing together with yComp the current graph and the rewrite process. This way you can see what the graph looks like at a given step of a complex transformation and develop the next step accordingly. Or you can debug your rules and sequences.
\end{description}

\section{The Core Assignment}

The task of the core assignment is to extract a state machine model out of the abstract syntax graph of a Java program.
The task stems from the domain of reengineering where software engineers need to gain insights into legacy systems, which is a lot easier given a birds eye view on the high level structure (and thus behavior) of the program.
The aim of the task is to evaluate the solutions and the tools backing them regarding performance and scalability, with a domain leading naturally to large graphs to be considered;
and especially to evaluate the solutions and tools regarding the ability to -- and conciseness in -- carrying out complex, non-local matchings of graph elements, requiring the matching of recursive graph structures.
Before the extraction can take place, the Java program graph needs to be imported from an Ecore file describing the source metamodel and an XMI file specifying the graph.
Afterwards the resulting state machine has to be exported into an XMI file conforming to a given Ecore file describing the state machine metamodel.

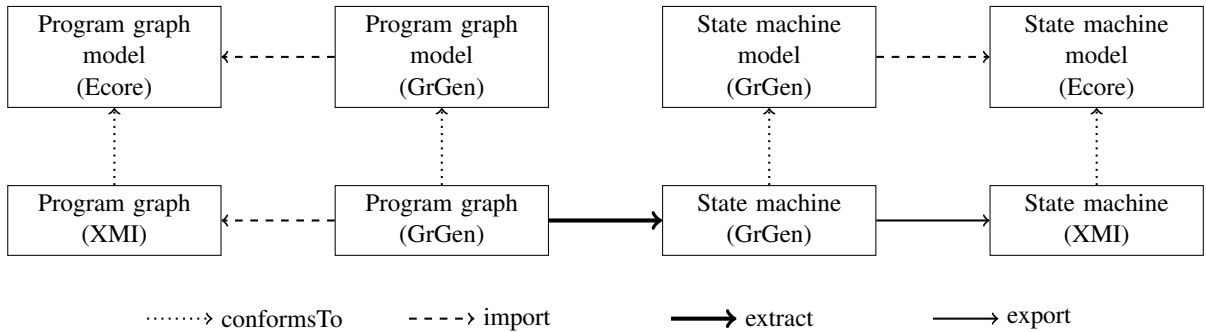
\begin{figure}[tb]
	\resizebox{\textwidth}{!}{\input{overview.tex}}
	\caption{Processing steps of the model extraction. The extraction and the XMI export are written in GrGen.NET languages. Import is handled by a supplied import filter, which generates GrGen meta model files (\texttt{.gm}) as an intermediate step.}
	\label{fig:overview}
\end{figure}

\subsection{Importing the Graph}

As GrGen.NET is a general purpose graph rewrite system and not a model transformation tool,
we do not support importing Ecore metamodels directly.
Instead, we supply an import filter generating an equivalent GrGen-specific graph model (.gm file) from a given Ecore file
by mapping classes to GrGen node classes,
their attributes to corresponding GrGen attributes,
and their references to GrGen edge classes.
Inheritance is transferred one-to-one, and enumerations are mapped to GrGen enums.
Class names are prefixed by the names of the packages they are contained in to prevent name clashes;
the same holds for references which are prefixed by their node class name.
Afterwards the instance graph XMI adhering to the metamodel described in the Ecore file thus adhering to the just generated equivalent GrGen graph model is imported by the filter into the system to serve as the host graph for the following extractions, i.e.\ transformations.
The entire process is shown in \autoref{fig:overview} above.

\subsection{Extracting the States}

The transformation is done in two steps, the first creates the states of the state machine, and the second inserts the transitions in between.
Each step consists of the application of one rule (utilizing a subpattern) on all matches found, giving a direct correspondence between coding conventions and rules.

Let us start with a short introduction into the syntax of the basic constructs of the rule language:
Rules in GrGen consist of a pattern part specifying the graph pattern to match and a nested rewrite part specifying the changes to be made.
The pattern part is built up of node and edge declarations or references with an intuitive syntax:
Nodes are declared by \texttt{n:t}, where \texttt{n} is an optional node identifier, and \texttt{t} its type.
An edge \texttt{e} with source \texttt{x} and target \texttt{y} is declared by \texttt{x -e:t-> y}, whereas \texttt{-->} introduces an anonymous edge of type \texttt{Edge}.
Nodes and edges are referenced outside their declaration by \texttt{n} and \texttt{-e->}, respectively.
Attribute conditions can be given within \texttt{if}-clauses.

The rewrite part is specified by a \texttt{replace} or \texttt{modify} block nested within the rule.
With \texttt{replace}-mode, graph elements which are referenced within the replace-block are kept, graph elements declared in the replace-block are created, and graph elements declared in the pattern, not referenced in the replace-part are deleted.
With \texttt{modify}-mode, all graph elements are kept, unless they are specified to be deleted within a \texttt{delete()}-statement.
Attribute recalculations can be given within an \texttt{eval}-statement.
These and the language elements we introduce later on are described in more detail in our solution of the Hello World case \cite{hello},
and especially in the extensive GrGen.NET user manual \cite{GrGenUserManual}.

\noindent Now let us have a look at the code to create the states (here and in the following rules the prefixes from name mangling were removed due to space constraints):

\begin{lstlisting}[language=grgen]
rule createStates
{
  stateClass:Class;
  stateClass -:annotationsAndModifiers-> :Abstract;
  if { stateClass.name == "State"; }

  es:CreateStates(stateClass);

  modify {
    sm:StateMachine;
    es(sm);
  }
}
\end{lstlisting}

We search for the abstract class of name \texttt{State} as starting point and create the state machine which will receive the states and transitions found.
The real work is done in a subpattern \texttt{CreateStates}, of which an instance \texttt{es} is declared and thus searched from the found \texttt{stateClass} on; or better in the rewrite part of this subpattern, which is applied with rule call syntax passing the just created \texttt{State\-Machine} node:

\begin{lstlisting}[language=grgen]
pattern CreateStates(parentClass:Class) modify(sm:StateMachine)
{
  iterated {
    extendingClass:Class -:extends-> r:NamespaceClassifierReference;
    r -:classifierReferences-> cr:ClassifierReference;
    cr -:target-> parentClass;

    es:CreateStates(extendingClass);

    optional {
      negative {
        extendingClass -:annotationsAndModifiers-> :Abstract;
      }

      modify {
        sm -:states-> s:State;
        s -:link-> extendingClass;
        eval { s.name = extendingClass.name; }
      }
    }

    modify {
      es(sm);
    }
  }

  modify { }
}
\end{lstlisting}

The subpattern searches a class directly extending the given parent class passed as a parameter,
i.e.\ it descends one subtyping step downwards in the type hierarchy.
It matches into breadth with the \texttt{iterated} construct to get all such extending classes;
the \texttt{iterated} matches all instances of its contained pattern which can be found in the host graph.
Then the subpattern matches further into depth by calling itself recursively with the just matched extending class as parameter, this way we cover the entire type hierarchy from the parent class on.
In the optional case the class is not abstract a state is created within the state machine and a \texttt{link} edge is created linking the state with the class (the \texttt{link} edges are additional helper edges introduced to store the traceability relation between source and target nodes).
The \texttt{optional} matches the contained pattern if it is available in the host graph.
The \texttt{negative} causes matching of the containing pattern to fail if its pattern can be found in the host graph.

\subsection{Extracting the Transitions}
The transitions are inserted with a second rule given in \autoref{createtransitionrule.grg} utilizing the subpattern given in \autoref{createtransitionpattern.grg}.
We search for the \texttt{class.Instance().activate()} pattern in the graph, if found we know the target state from the class of the called method and the link between the class and the state we inserted previously.
Then we search with the subpattern \texttt{FindSourceState} for the source state, which gets \texttt{yield}ed
\footnote{The \texttt{yield}-keyword is used to open a block with yielding assignments, and as a mandatory annotation at an entity whenever it is yielded to.}
into the \texttt{def} pattern element \texttt{sourceState}.
If this was found we add a \texttt{Transition} in between the source state and the target state,
and link it to the \texttt{expressionStatement} containing the method call.

The subpattern is used to recursively walk outwards from the method call to the class containing the call (recursive ascent over statement nesting); passing over the different types of statements and statement containers which might be on the way, until the class is reached yielding it back.
The statements passed are all \texttt {link}ed to the transition, which will be helpful for the extension tasks.

\subsection{Extension Tasks}
The trigger attributes of the transitions are filled by four rules for the four different ways specified;
they get executed one after the other (this way handling the priority), first the non-run method, then the switch case, then the catch block and finally the fallback rule.
Due to the links from the transitions to all the constructs on the path from the method call to the containing class this is a simple local pattern search, as can be seen in \autoref{extension1.grg}.

The action attributes of the transitions are filled by two rules for the two different ways specified;
first the enum value used in a send method, then the fallback rule.
Again this is a simple local pattern search due to the links from the transitions to all the constructs on the path from the method call to the containing class, cf. \autoref{extension2.grg}.

\subsection{Exporting the State Machine}
A visualization of the resulting state machine is given in \autoref{fig:statemachine}.
The XMI export of this graph is handled by 5 additional rules given in \texttt{export.gri} containing \texttt{emit} statements:
one for assigning XMI ids to the elements to be exported, which are stored in a map from the nodes to the corresponding ids, two for writing the XMI prefix and suffix, one for writing the \texttt{States} and one for writing the \texttt{Transitions}, utilizing the previously computed node to id mapping.

\section{Rule Control, Performance, and Visualization}

The extraction process is controlled by the graph rewrite script \texttt{reengineering.grs} which is executed by the GrShell;
the script contains the following lines:

\begin{lstlisting}[language=grshell]
import primitive_types.ecore java.ecore StateMachine.ecore 
         1_small-model.xmi reengineering.grg

xgrs [createStates]
xgrs [createTransitions] 
xgrs [addTriggerNonRunMethodName] ;> [addTriggerSwitchCaseEnumValueName] \
       ;> [addTriggerCatchBlockExceptionClassName] ;> [addTriggerOtherwise]
xgrs [addActionSend] ;> [addActionOtherwise]
\end{lstlisting}

The import command imports the XMI input graph complying to the Ecore models, and additionally includes the rules given in the rule file.
The \texttt{xgrs} keyword starts an extended graph rewrite sequence, which is the rule application control language of GrGen
(prepending \texttt{debug} before \texttt{xgrs} allows you to debug the sequence execution in GrShell).
The rules are executed on all the matches found, which is requested by the all-bracketing \texttt{[rule]}.
The then-right operator \texttt{;>} executes the left sequences, then the right sequence, and returns as sequence result the result of the right sequence;
the sequence results are irrelevant for this task, in general they are used to control sequence applications.

\subsection{Performance}

The benchmark results for the extraction task are given in the following table.

\begin{table}[h]
\centering
\label{tab:ne}
\begin{tabular}{r||r|r|r|r|r}
	set no. & import time & import size & shell time & shell size & extraction time \\ \hline
       1 &      2,855 &           2.0 &           31 &     3.5 &          130\\
       2 &      2,917 &           2.1 &           32 &     3.6 &          140\\
       3 &     17,878 &         188.8 &        4,165 &   420.9 &          187\\ \hline
       1 &      1,279 &           1.3 &           46 &     2.3 &          125\\
       2 &      1,314 &           1.3 &           47 &     2.4 &          130\\
       3 &     28,658 &         105.3 &        7,800 &   277.4 &          213\\
\end{tabular}
\caption{Results for different input sets; running time in ms, memory usage in MiBytes.}
\end{table}

\noindent The reported values are computed as the arithmetic mean of the middle 3 values out of 5 measurements, on a Core i7 920 (2.6GHz) with 6 GiBytes of main memory under Windows Vista 64~Bit with MS .NET 64~Bit for the upper part of the table and on a Core 2 Duo U9600 (1.6GHz) with 3 GiBytes of main memory under Windows 7 32~Bit with MS .NET 32~Bit for the lower part.
Import time is the time needed for importing the graph, import size is the size of the heap after importing the graph.
Shell time is the time needed to transform the imported graph from API level (as seen by an external program using the API) into a named graph as employed by the GrShell of the rapid prototyping environment, shell size is the size of the heap after the named graph was constructed.
Extraction time is the time needed for the application of the extraction rules.
Remark: the dominating component of the extraction time is the time needed by the .NET just-in-time compiler producing machine code out of the .NET bytecode.

\subsection{Visualization}
The GrShell utilizes the graph viewer yComp as its visualization component; the final state machine visualization was already presented in \autoref{fig:statemachine}.
But in addition to the state machine, yComp is able to give a decent visualization of the original program graph, too, as you may see in Figures~\ref{fig:programgraph}, \ref{fig:programgraphzoom} and~\ref{fig:programgraphmorezoom} which give a series of images zoomed in, an outstanding help in understanding and debugging.

This is made possible by the high configurability of yComp:
one can choose from several available layout algorithms, e.g. organic for a force-based layout, or hierarchic which works well for program graphs (it was used in rendering the syntax graphs shown in this paper).
For every available node or edge type it can be configured in which color with what node shape or edge style it should be shown, with what attribute values or fixed text as element labels or tags it is to be displayed, or if it should be shown at all.
Furthermore graph nesting can be configured by registering edges at certain nodes to define a containment hierarchy, causing the nodes to be displayed as subgraphs containing the elements to which they are linked by the given edges.
In \autoref{layout.grsi} an excerpt from our configuration file for customizing the graph layout of the program graph is shown.
In addition a helper edge introduction step was added, so that all expression nodes are nested inside their containing statements, not only the outermost ones.
A helper step was used in producing the final state machine visualization \autoref{fig:statemachine}, too, replacing Transition nodes with real edges.

\section{Conclusion}
In this paper we presented a GrGen.NET solution to the Reengineering challenge of the Transformation Tool Contest 2011 (which is available as a SHARE image~\cite{share}).
The abstract Java syntax graph conforming to the \texttt{java.ecore} metamodel was imported by a supplied import filter under remapping to the graph concepts supported by GrGen; the extracted state machine was exported by a handful of text emitting graph rewrite rules.
The name mangling Ecore/XMI import and the explicitely programmed XMI export, which result from the fact that GrGen is not a dedicated model transformation tool, are the only points of our solution we regard to be rather weak.
In contrast to the declarative graph rewriting with recursive and iterated structures (see \cite{StructuralRecursion} for more on this) which we employed to extract a state machine giving a high level overview out of the original syntax graph.
This ability of matching and rewriting recursive patterns allowed us to give a concise and simple solution to the core task of the Reengineering challenge closely following the specification given,
with rules matching kernel patterns and subpattern recursion and iteration to match recursive structures into depth and breadth.
During rewriting of the recursive match for transition creation, from the activation call to the containing class outwards, links were inserted from the transition to the elements visited;
besides having been a help in debugging they especially allowed to easily solve the extension task with purely local graph rewrite rules (we could have used storagemaps instead of storing and retrieving this tracebility information as other tools have done, but using edges allows for easy, visual debugging).
The extraction resulted in correct target state machines for all input graphs supplied;
for the large graph it took about 200ms, i.e.\ performance is an order of magnitude better than the reference solution.
The goal of the task is to allow program understanding by extracting and displaying a reduced, easily understandable model.
In addition to visualizing this simple model we have presented a visualization of the original program graph with our graph viewer yComp; utilizing color customization and especially nesting subgraphs inside nodes, we managed to achieve an understandable visualization for the two medium sized example graphs (7000 nodes plus 7000 edges), which a reengineer can inspect and navigate in order to develop and debug his transformation rules.


\nocite{*}
\bibliographystyle{eptcs}
\bibliography{refs}

\vspace{1cm}
\appendix
\section{Code Listings and Screenshots}
\input{appendix.tex}

\end{document}

%% file: overview.tex
\begin{tikzpicture}[scale=0.9, transform shape]
\path 
	  (0,0)     node[block] (xmi14)   {Program graph\\(XMI)}
	 +(0,2.5)   node[block] (ecore14)   {Program graph\\ model\\(Ecore)}
	 ++(5,0)    node[block] (grg14)  {Program graph\\(GrGen)}
	 +(0,2.5)   node[block] (gm14)   {Program graph\\ model\\(GrGen)}
	 ++(5,0)    node[block] (grg21) {State machine\\(GrGen)}
	 +(0,2.5)   node[block] (gm21)   {State machine\\ model\\(GrGen)}
	 ++(5,0)    node[block] (xmi21)   {State machine\\(XMI)}
	 +(0,2.5)   node[block] (ecore21)   {State machine\\ model\\(Ecore)}

	 (0.5,-1.5) coordinate  (leftConform)
	 +(1,0)     coordinate  (rightConform) node[right] {conformsTo}
	 ++(4,0)    coordinate  (leftImport)
	 +(1,0)     coordinate  (rightImport) node[right] {import}
	 ++(4,0)    coordinate  (leftTransform)
	 +(1,0)     coordinate  (rightTransform) node[right] {extract}
	 ++(4,0)    coordinate  (leftExport)
	 +(1,0)     coordinate  (rightExport) node[right] {export}
;

\draw[import] (gm14) -- (ecore14);
\draw[import] (gm21) -- (ecore21);
\draw[import] (grg14) -- (xmi14);
\draw[export] (grg21) -- (xmi21);

\draw[conform] (xmi14) -- (ecore14);
\draw[conform] (grg14) -- (gm14);
\draw[conform] (grg21) -- (gm21);
\draw[conform] (xmi21) -- (ecore21);

\draw[transform] (grg14) -- (grg21);

\draw[conform] (leftConform) -- (rightConform);
\draw[import] (leftImport) -- (rightImport);
\draw[export] (leftExport) -- (rightExport);
\draw[transform] (leftTransform) -- (rightTransform);
\end{tikzpicture}

%% file: appendix.tex
\begin{figure}[hb]
\begin{lstlisting}[language=grgen]
rule createTransitions
{
  expressionStatement:ExpressionStatement -:expression-> refTargetClass;
  refTargetClass:IdentifierReference -:target-> targetClass:Class;
  refTargetClass -:next-> callInstance;
  callInstance:MethodCall -:target-> instance:ClassMethod;
  callInstance -:next-> callActivate;
  callActivate:MethodCall -:target-> activate:ClassMethod;
  if { instance.name=="Instance" && activate.name=="activate"; }

  targetClass <-:link- targetState:State;
  def sourceState:State;
  fss:FindSourceState(expressionStatement, yield sourceState);

  sm:StateMachine;

  modify {
    sm -:transitions-> transition:Transition;
    sourceState <-:src- transition -:dst-> targetState;
    sourceState -:out-> transition <-:in- targetState;
    transition -:link-> expressionStatement;
    fss(transition);
  }
}
\end{lstlisting}
	\caption{Rule for creating transitions}
	\label{createtransitionrule.grg}
\end{figure}

\begin{figure}
\begin{lstlisting}[language=grgen]
pattern FindSourceState(containedEntity:Node, def sourceState:State)
                        modify(transition:Transition)
{
  alternative {
    StatementListContainer {
      listContainer:StatementListContainer -:statements-> containedEntity;
      fss:FindSourceState(listContainer, yield sourceState);
      modify {
        transition -:link-> listContainer;
        fss(transition);
      }
    }
    StatementContainer {
      container:StatementContainer -:statement-> containedEntity;
      fss:FindSourceState(container, yield sourceState);
      modify {
        transition -:link-> container;
        fss(transition);
      }
    }
    StatementSwitch {
      switch:Switch -:cases-> containedEntity;
      fss:FindSourceState(switch, yield sourceState);
      modify {
        transition -:link-> switch;
        fss(transition);
      }
    }
    StatementCondition {
      condition:Condition -:elseStatement-> containedEntity;
      fss:FindSourceState(condition, yield sourceState);
      modify {
        transition -:link-> condition;
        fss(transition);
      }
    }
    StatementTry {
      try:TryBlock -:catcheBlocks-> containedEntity;
      fss:FindSourceState(try, yield sourceState);
      modify {
        transition -:link-> try;
        fss(transition);
      }
    }
    Class {
      cc:Class -:members-> containedEntity;
      ss:State -:link-> cc;
      yield { yield sourceState = ss; }
      modify {
        transition -:link-> cc;
      }
    }
  }
}
\end{lstlisting}
	\caption{Subpattern for creating transitions}
	\label{createtransitionpattern.grg}
\end{figure}

\begin{figure}
\begin{lstlisting}[language=grgen]
rule addTriggerNonRunMethodName
{
  transition:Transition -:link-> method:ClassMethod;
  if { method.name != "run"; }

  modify {
    eval { transition.trigger = method.name; }
  }
}

rule addTriggerSwitchCaseEnumValueName
{
  transition:Transition -:link-> case:NormalSwitchCase;
  case -:condition-> caseCondition:IdentifierReference;
  caseCondition -:target-> value:EnumConstant;

  modify {
    eval { transition.trigger = value.name; }
  }
}

rule addTriggerCatchBlockExceptionClassName
{
  transition:Transition -:link-> catchBlock:CatchBlock;
  catchBlock -:parameter-> parameter:OrdinaryParameter;
  parameter -:typeReference-> nspClassRef:NamespaceClassifierReference;
  nspClassRef -:classifierReferences-> classRef:ClassifierReference;
  classRef -:target-> exceptionClass:Class;

  modify {
    eval { transition.trigger = exceptionClass.name; }
  }
}

rule addTriggerOtherwise
{
  transition:Transition;
  if { transition.trigger == null || transition.trigger == ""; }

  modify {
    eval { transition.trigger = "--"; }
  }
}
\end{lstlisting}
	\caption{Rules for extension task 1}
	\label{extension1.grg}
\end{figure}

\begin{figure}
\begin{lstlisting}[language=grgen]
rule addActionSend
{
  transition:Transition -:link-> block:StatementListContainer;
  block -:statements-> exprStmt:ExpressionStatement;
  exprStmt -:expression-> callMethod:MethodCall;
  callMethod -:target-> method:ClassMethod;
  callMethod -:arguments-> enumClassRef:IdentifierReference;
  enumClassRef -:next-> enumValueRef:IdentifierReference;
  enumValueRef -:target-> enumValue:EnumConstant;
  if { method.name == "send"; }

  modify {
    eval { transition.action = enumValue.name; }
  }
}

rule addActionOtherwise
{
  transition:Transition;
  if { transition.action == null || transition.action == ""; }

  modify {
    eval { transition.action = "--"; }
  }
}
\end{lstlisting}
	\caption{Rules for extension task 2}
	\label{extension2.grg}
\end{figure}

\begin{figure}
\centering
\includegraphics[scale=0.82]{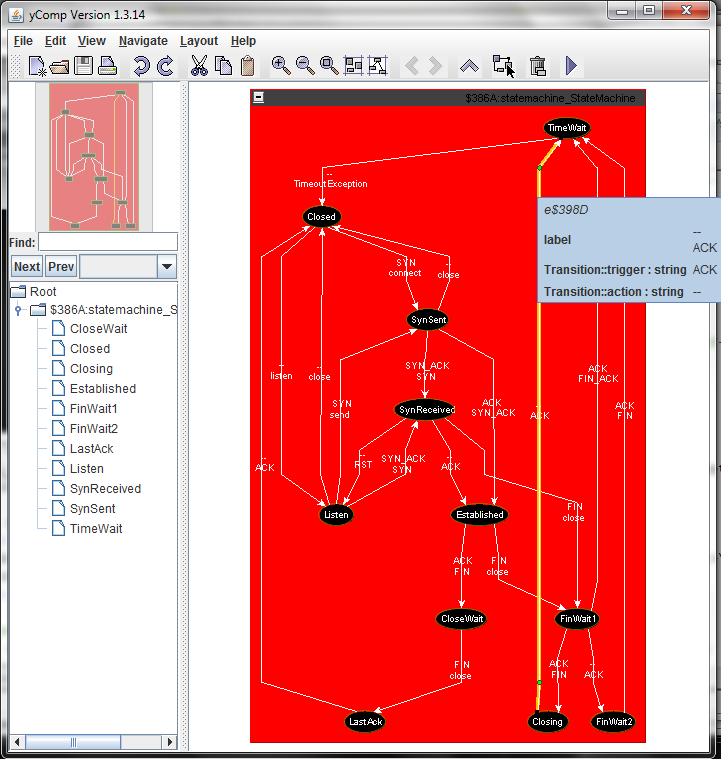}
\caption{The resulting state machine, with an edge selected and its attributes displayed}
\label{fig:statemachine}
\end{figure}

\begin{figure}
\centering
\includegraphics[scale=0.27]{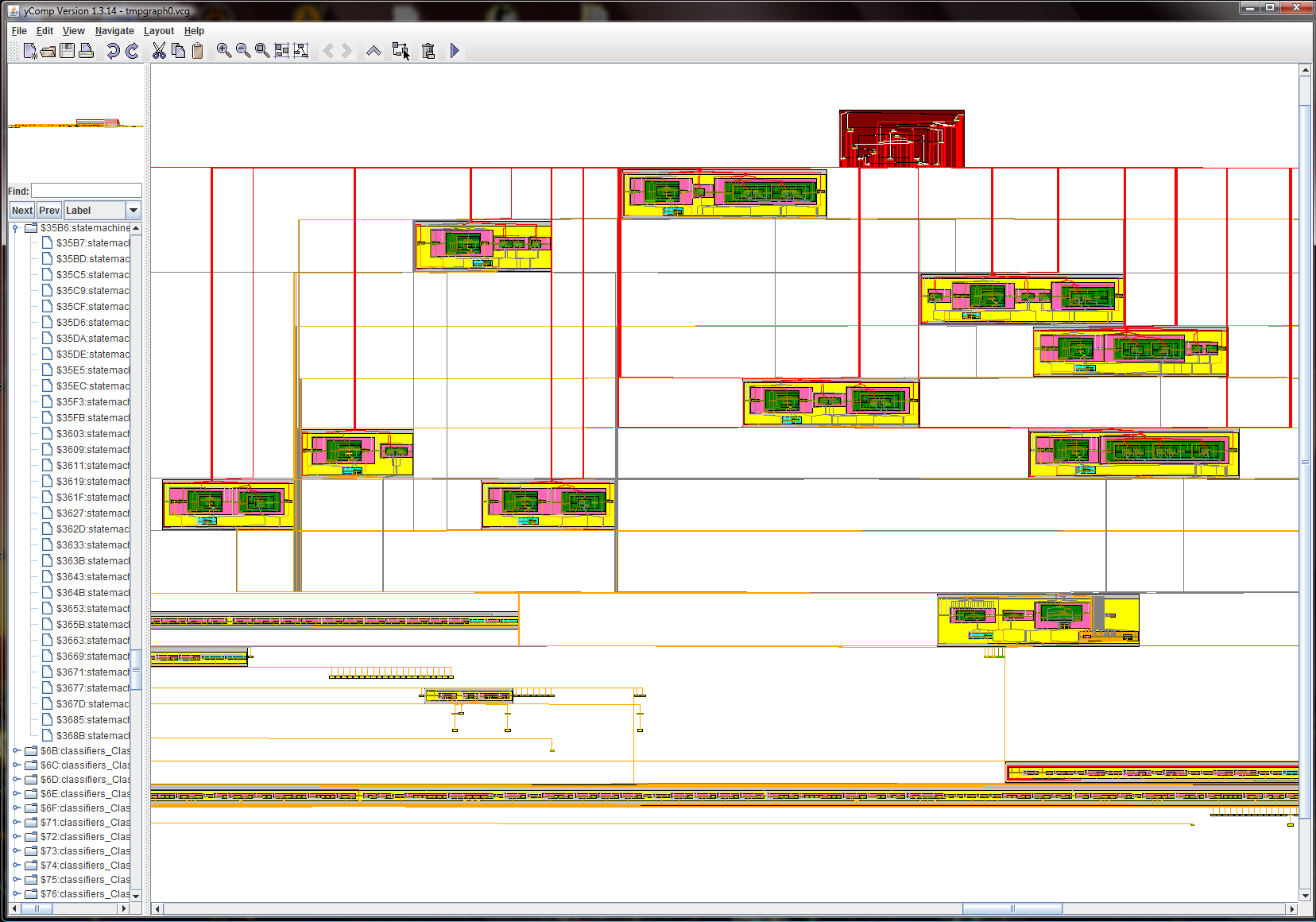}
\caption{The program graph, the rectangles are the classes, the top one is the state machine}
\label{fig:programgraph}
\end{figure}

\begin{figure}
\centering
\includegraphics[scale=0.27]{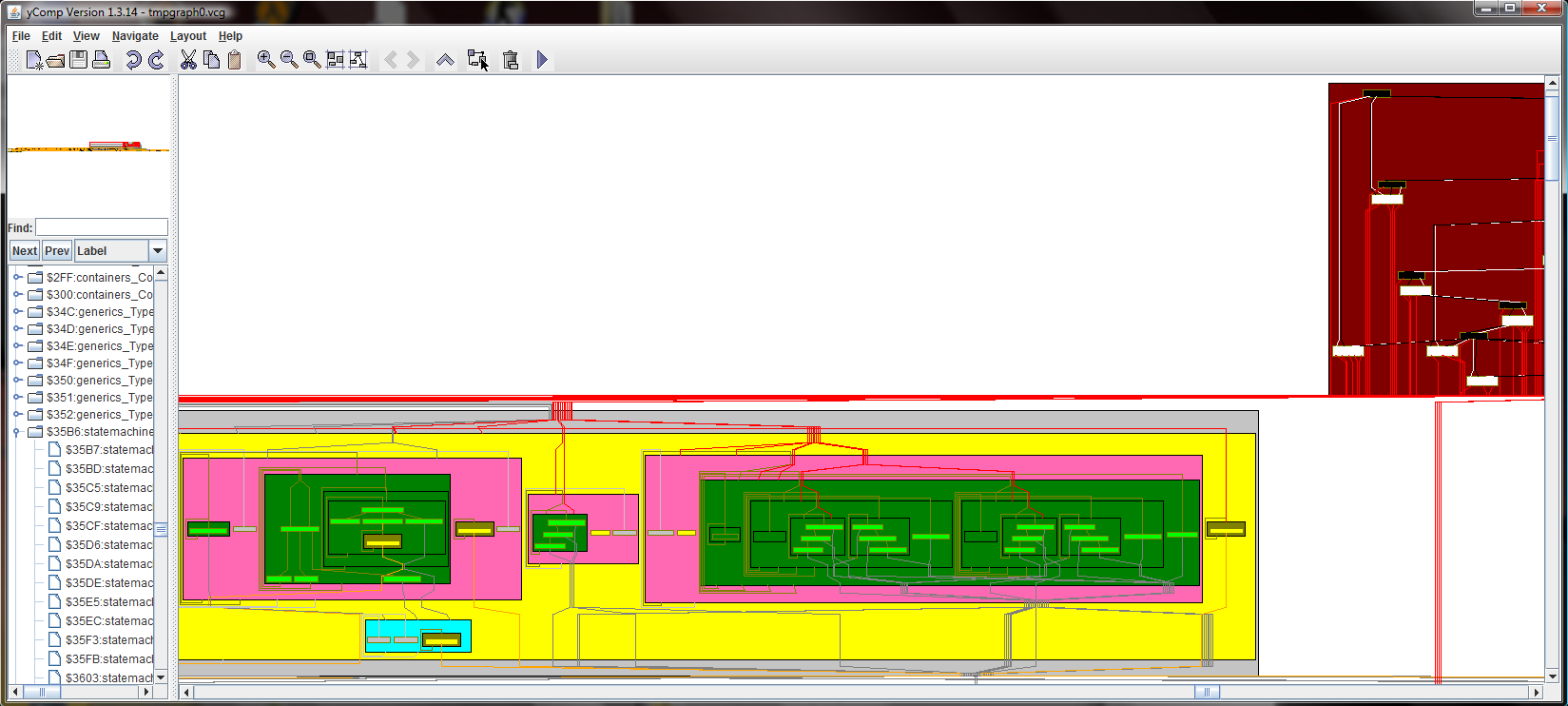}
\caption{The program graph zoomed with the class SynSent and the state machine}
\label{fig:programgraphzoom}
\end{figure}

\begin{figure}
\centering
\includegraphics[scale=0.27]{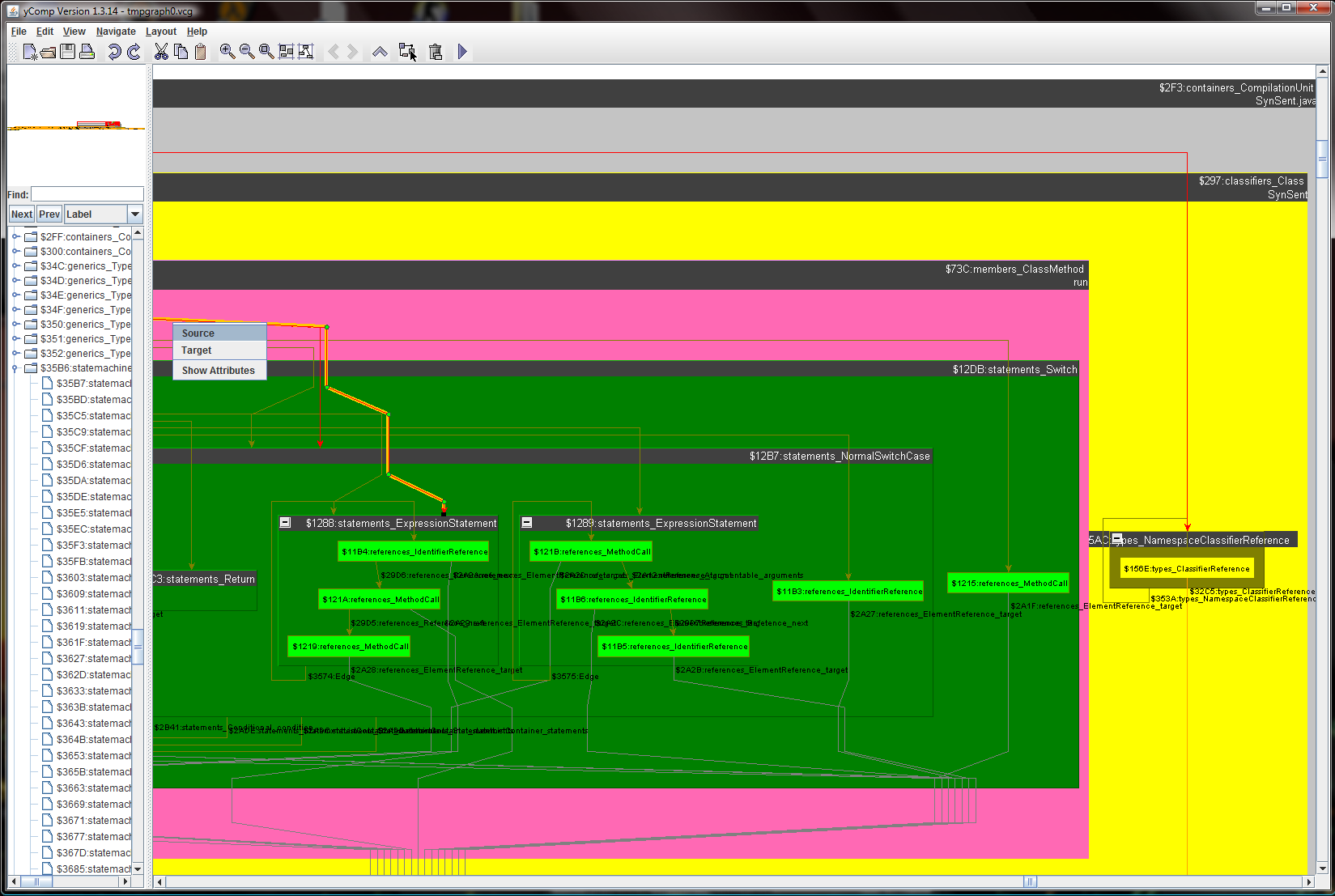}
\caption{The program graph zoomed further to the method run of the class SynSent}
\label{fig:programgraphmorezoom}
\end{figure}

\begin{figure}
\begin{lstlisting}[language=grshell]
debug set layout Hierarchic

dump add node classifiers_Classifier
       group by hidden outgoing members_MemberContainer_members
dump add node members_Method
       group by hidden outgoing members_MemberContainer_members

dump add node classifiers_Annotation exclude

dump set node members_Method color pink
dump set node members_Field color cyan

dump set edge references_ElementReference_target color grey

dump add node classifiers_Class shortinfotag _name
dump add node members_Method shortinfotag _name
\end{lstlisting}
	\caption{An excerpt from the configuration of the layout}
	\label{layout.grsi}
\end{figure}